\documentclass[aps,prb,twocolumn,showpacs,preprintnumbers,amsmath,amssymb,superscriptaddress]{revtex4}%

\usepackage{graphicx}%
\usepackage{dcolumn}
\usepackage{amsmath}
\usepackage{color}

\begin{document}

\preprint{PREPRINT (\today)}

\title{Nodeless superconductivity in the infinite-layer electron-doped
Sr$_{0.9}$La$_{0.1}$CuO$_2$ cuprate superconductor}

\author{R.~Khasanov}
 \email{rustem.khasanov@psi.ch}
 \affiliation{Laboratory for Muon Spin Spectroscopy, Paul Scherrer
Institut, CH-5232 Villigen PSI, Switzerland}
\author{A.~Shengelaya}
 \affiliation{Physics Institute of Tbilisi State University,
Chavchavadze 3, GE-0128 Tbilisi, Georgia}
\author{A.~Maisuradze}
 \affiliation{Physik-Institut der Universit\"{a}t Z\"{u}rich,
Winterthurerstrasse 190, CH-8057 Z\"urich, Switzerland}
\author{D.~Di~Castro}
 \affiliation{Physik-Institut der Universit\"{a}t Z\"{u}rich,
Winterthurerstrasse 190, CH-8057 Z\"urich, Switzerland}
 \affiliation{"Coherentia" CNR-INFM and Dipartimento di Fisica,
Universita' di Roma "La Sapienza", P.le A. Moro 2, I-00185 Roma,
Italy}
\author{I.M.~Savi\'c}
 \affiliation{Faculty of Physics, University of Belgrade, 11001
Belgrade, Serbia and Montenegro}
\author{S.~Weyeneth}
 \affiliation{Physik-Institut der Universit\"{a}t Z\"{u}rich,
Winterthurerstrasse 190, CH-8057 Z\"urich, Switzerland}
\author{M.S.~Park}
 \affiliation{National Creative Research Initiative Center for
Superconductivity and Department of Physics, Pohang University of
Science and Technology, Pohang 790-784, Republic of Korea}
\author{D.J.~Jang}
 \affiliation{National Creative Research Initiative Center for
Superconductivity and Department of Physics, Pohang University of
Science and Technology, Pohang 790-784, Republic of Korea}
\author{S.-I.~Lee}
 \affiliation{National Creative Research Initiative Center for
Superconductivity and Department of Physics, Pohang University of
Science and Technology, Pohang 790-784, Republic of Korea}
 \affiliation{Department of Physics, Sogang University, Seoul 121-742, Korea}
\author{H.~Keller}
 \affiliation{Physik-Institut der Universit\"{a}t Z\"{u}rich,
Winterthurerstrasse 190, CH-8057 Z\"urich, Switzerland}

\begin{abstract}
We report on  measurements of the in-plane magnetic  penetration
depth $\lambda_{ab}$ in the infinite-layer electron-doped
high-temperature cuprate superconductor
Sr$_{0.9}$La$_{0.1}$CuO$_{2}$ by means of muon-spin rotation.
The observed temperature and magnetic field dependences of
$\lambda_{ab}$ are consistent with the presence of a substantial
$s-$wave component in the superconducting order parameter in good
agreement with the results of tunneling, specific heat, and
small-angle neutron scattering experiments.
\end{abstract}

\maketitle

\section{Introduction}

The symmetry of the superconducting energy gap is one of the
essential issues for understanding  the mechanism of
high-temperature superconductivity. For hole-doped
high-temperature cuprate superconductors (HTS's) it is commonly
accepted that the superconducting energy gap  has $d-$wave
symmetry, as indicated, {\it e.g.}, by tricrystal
experiments,\cite{Tsuei94} although a  multi-component
($d+s-$wave) gap is now acquiring overwhelming evidence.
\cite{Khasanov07_La214,Khasanov07_Y124,Khasanov07_Y123} For
electron-doped HTS's, however, no consensus was reached on this
issue so far. A number of experiments, including angular resolved
photoemission,\cite{Sato01,Matsui05} scanning-SQUID,\cite{Tsuei00}
Raman scattering,\cite{Blumberg02} and magnetic penetration depth
studies\cite{Kokales00,Snezhko04} point to a $d-$wave, or more
general, to a nonmonotonic $d-$wave gap (with the gap maximum in
between the nodal and the antinodal points on the Fermi surface)
in electron-doped Nd$_{2-x}$Ce$_x$CuO$_4$ and
Pr$_{2-x}$Ce$_x$CuO$_4$. On the contrary, a state corresponding to
an $s-$ or a nonmonotonic $s-$wave gap symmetry was reported for
similar compounds and infinite-layer Sr$_{1-x}$La$_{x}$CuO$_{2}$
in tunnelling,\cite{Shan05,Chen02} Raman
scattering,\cite{Hackl95-03} penetration depth,\cite{Alff99}
small-angle neutron scattering,\cite{White08} and specific heat
\cite{Liu05} studies. Results of Biswas {\it et al.}
\cite{Biswas02} and Skinta {\it et al.}\cite{Skinta02} suggest
that there is a $d-$ to $s-$wave transition across optimal doping
in Nd$_{2-x}$Ce$_x$CuO$_4$ and Pr$_{2-x}$Ce$_x$CuO$_4$. The
two-gap picture was also introduced in Refs.~\onlinecite{Luo05}
and \onlinecite{Liu06} in order to explain the unusual behavior of
the magnetic penetration depth and Raman spectra.

Here we report a study of the in-plane magnetic field penetration
depth $\lambda_{ab}$ in the infinite-layer electron-doped
superconductor Sr$_{0.9}$La$_{0.1}$CuO$_{2}$ by means of
transverse-field muon-spin rotation (TF-$\mu$SR). This compound
belongs to the family of electron-doped HTS's (Sr,Ln)CuO$_2$
(Ln=La, Sm, Nd, Gd) with the so-called infinite-layer
structure.\cite{Siegrist88,Smith91} It has the simplest crystal
structure among all HTS's consisting of an infinite stacking of
CuO$_2$ planes and (Sr, Ln) layers. The charge reservoir block,
commonly present in HTS's, does not exist in the infinite-layer
structure. It also has stoichiometric oxygen content without
vacancies or interstitial oxygen,\cite{Jorgensen93} which is a
general problem for most of the electron- and the hole-doped
HTS's. In the present study $\lambda_{ab}^{-2}(T)$ was
reconstructed from the measured temperature dependences of the
$\mu$SR linewidth by applying the numerical calculations of
Brandt.\cite{Brandt03} $\lambda_{ab}$ was found to be almost field
independent in contrast to the strong magnetic field dependence
observed in the hole-doped HTS's, suggesting that there are no
nodes in the superconducting energy gap of
Sr$_{0.9}$La$_{0.1}$CuO$_{2}$. The temperature dependence of
$\lambda_{ab}^{-2}$ was found to be well described by anisotropic
$s-$wave as well as by two-gap models ($d+s$ and $s+s$). In the
case of the two-gap $d+s$ model the contribution of the $d-$wave
gap to the total superfluid density was found to be of the order
of 15\%. Our results imply that a substantial $s-$wave component
in the superconducting order parameter is present in
Sr$_{0.9}$La$_{0.1}$CuO$_{2}$, in agreement with previously
reported results of tunnelling,\cite{Chen02} specific
heat,\cite{Liu05} and small-angle neutron scattering
experiments.\cite{White08}
%

\section{Experimental details}

Details on the sample preparation for
Sr$_{0.9}$La$_{0.1}$CuO$_{2}$ can  be found
elsewhere.\cite{Jung02}
The TF-$\mu$SR experiments were performed at  the
$\pi$M3 beam line at the Paul Scherrer Institute (Villigen,
Switzerland). The sintered Sr$_{0.9}$La$_{0.1}$CuO$_{2}$ sample
was field cooled from above $T_c$ to 1.6~K in a series of fields
ranging from 50~mT to 0.64~T.

In the transverse-field geometry the local magnetic field
distribution $P(B)$ inside the superconducting sample in the mixed
state, probed by means of TF-$\mu$SR, is determined
by the values of the coherence length $\xi$ and the magnetic field
penetration depth $\lambda$. In extreme type-II superconductors
($\lambda\gg\xi$) $P(B)$ is almost independent on
$\xi$ and the second moment of $P(B)$ becomes proportional to
$1/\lambda^4$.\cite{Brandt88,Brandt03}
In order to describe the asymmetric local magnetic field
distribution $P(B)$ in the superconductor in the mixed state, the
analysis of the data was based on a two component Gaussian fit of
the $\mu$SR time spectra:\cite{Khasanov06a}
\begin{equation}
P(t)= \sum_{i=1}^2A_i \exp(-\sigma_i^2t^2/2)
\cos(\gamma_{\mu}B_i t+\phi).
\label{eq:gauss}
\end{equation}
Here $A_i$, $\sigma_i$, and $B_i$ are the asymmetry, the
relaxation rate, and the mean field of the $i-$th component,
$\gamma_\mu = 2\pi\times135.5342$~MHz/T denotes the muon
gyromagnetic ratio, and $\phi$ is the initial phase of the
muon-spin ensemble.
The total second moment of the $\mu$SR line was derived
as:\cite{Khasanov06a}
\begin{equation}
\langle \Delta B^{2}\rangle=\frac{\sigma^2}{\gamma^2_\mu}=
\sum_{i=1}^2{A_i \over A_1+A_2} \left[
\frac{\sigma_i^2}{\gamma^2_\mu} +\left(B_i- \frac{A_i
B_i}{A_1+A_2}\right)^2 \right] .
\label{eq:dB}
\end{equation}
The superconducting part of the square root of the  second moment
$\sigma_{sc}$ was further obtained by subtracting the contribution
of the nuclear moments $\sigma_{nm}$ measured at $T>T_c$ as
$\sigma_{sc}^2=\sigma^2 - \sigma_{nm}^2$.\cite{Zimmermann95}

The following issue is important for the interpretation  of the
experimental data: The sample used in the experiment was a
nonoriented sintered powder. In this case an effective averaged
penetration depth $\lambda_{eff}$ can be extracted. However, in
highly anisotropic extreme type-II superconductors (as HTS's)
$\lambda_{eff}$ is dominated by the in-plane penetration depth so
that $\lambda_{eff} \simeq1.31 \lambda_{ab}$.\cite{Fesenko91}

\section{Results and discussions}

\begin{figure}[htb]
\includegraphics[width=1.0\linewidth]{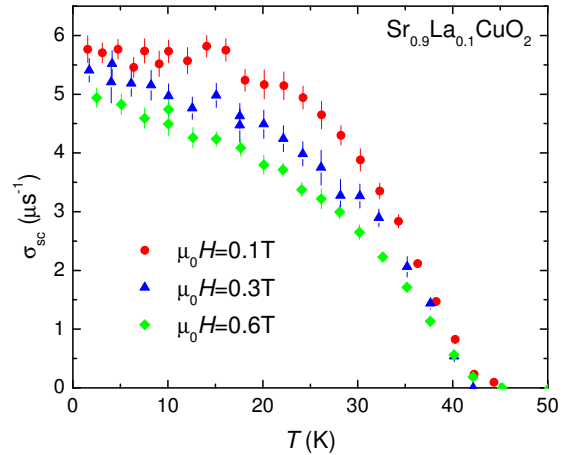}
\caption{(Color online) Temperature dependences of
$\sigma_{sc}\propto\lambda_{ab}^{-2}$ of
Sr$_{0.9}$La$_{0.1}$CuO$_{2}$ measured in magnetic fields of
0.1~T, 0.3~T, and 0.6~T.}
 \label{fig:sigma_vs_T}
\end{figure}

Figure~\ref{fig:sigma_vs_T} shows the  temperature  dependences of
$\sigma_{sc}\propto\lambda_{ab}^{-2}$  measured after
field-cooling the sample from far above $T_c$ in $\mu_0H$=0.1~T,
0.3~T, and 0.6~T.  Two features are clearly seen: (i) In the whole
temperature region (from $T\simeq1.6$~K up to $T_c$)
$\sigma_{sc}(T,H)$ decreases with increasing field. (ii) The
curvature of $\sigma_{sc}(T)$ changes with field. With decreasing
temperature $\sigma_{sc}$ at $\mu_0H=0.1$~T first increases and
then becomes $T$ independent for $T\lesssim15$~K, while
$\sigma_{sc}$ for both $\mu_0H=0.3$~T and $\mu_0H=0.6$~T increases
continuously in the whole range of temperatures. It is also seen
that the low-temperature slope of $\sigma_{sc}(T)$ is larger at
the higher fields.

The decrease of $\sigma_{sc}$ with increasing magnetic field is
caused by the overlapping of the vortices by their cores, leading
to a reduction of the field variance in the superconductor in the
mixed state. As shown by Brandt,\cite{Brandt03} at magnetic
inductions $B/B_{c2}\lesssim 0.1$ ($B_{c2}$ is the second critical
field), overlapping of vortex cores may be neglected and vortex
properties are independent of the applied magnetic field. Only the
vortex density is changed. At higher magnetic inductions vortices
start to overlap by their cores. Consequently, not only the vortex
density, but also the properties of the individual vortices become
magnetic field dependent.
Therefore, the different temperature behavior of $\sigma_{sc}$ can
be explained by the fact that the vortex core size, which is
generally assumed to be equal to the coherence length
$\xi=[\Phi_0/2\pi B_{c2}]^{0.5}$, increases with increasing
temperature (decreasing $B_{c2}$). Thus higher temperature [bigger
reduced field $b(T)=B/B_{c2}(T)$]  would correspond to larger
overlapping vortex cores leading to a stronger reduction of
$\sigma_{sc}$.  Consequently, the correcting factor between
$\sigma_{sc}$ and the magnetic field penetration depth is {\it not
constant}, as is generally assumed, but depends on the reduced
field $b$:\cite{Brandt03}
\begin{equation}
\sigma_{sc}(b)[\mu{\rm s}^{-1}]=A(b)\lambda^{-2}_{eff}[{\rm nm}^{-2}].
 \label{eq:sigma-lambda}
\end{equation}
Here $A(b)$ is a correcting factor, which for  superconductors
with a Ginzburg-Landau parameter $\kappa=\lambda/\xi\geq5$  in
the range of fields $0.25/\kappa^{1.3}\lesssim b\leq1$ can be
obtained analytically as $A(b)=4.83\cdot10^4 (1 - b)
 [1 + 1.21(1 - \sqrt{b})^3]$~$\mu$s$^{-1}$nm$^{2}$.\cite{Brandt03}

\begin{figure}[htb]
\includegraphics[width=0.8\linewidth]{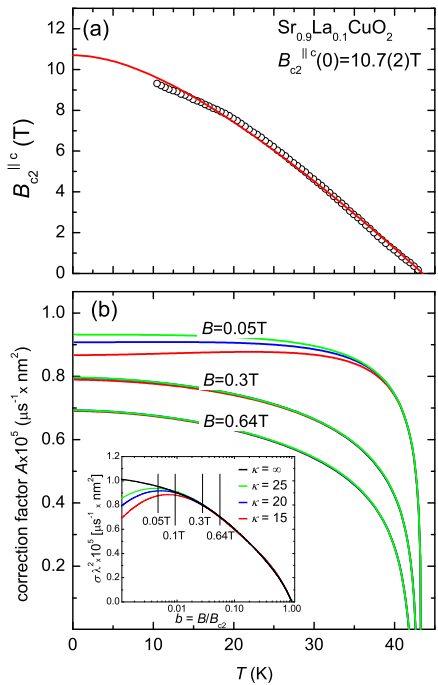}
\caption{(Color online)(a) Temperature dependence of  the second
critical field $B_{c2}^{\parallel c}(T)$ of
Sr$_{0.9}$La$_{0.1}$CuO$_{2}$ obtained from measurements of the
reversible magnetization on $c-$axis oriented powder by using the
Landau-Ott approach.\cite{Landau03} The solid line is the fit of
the Werthamer-Helfand-Hohenberg (WHH) model\cite{Werthamer66} to
the data with $B_{c2}^{\parallel c}(0)=10.7(2)$~T. (b) Temperature
dependence of the correcting factor $A$ at $B=0.05$~T, 0.1~T, and
$0.6$~T for $\kappa=\lambda/\xi=15$ (green line), 20 (blue line),
and 25 (red line). The inset shows $A(b)=\sigma\cdot\lambda^2$
obtained from numerical calculations using the model of
Brandt.\cite{Brandt03} Vertical lines mark the values of
$B/B_{c2}(0)$ for $B=0.05$~T, 0.1~T, 0.3~T, and 0.64~T.}
 \label{fig:Hc2_corrections}
\end{figure}

Equation~(\ref{eq:sigma-lambda}) requires that in order to derive
$\lambda$ from measured $\sigma_{sc}(T)$, the temperature
dependence of $B_{c2}$ must be taken into account. Since in our
experiments the measured $\lambda_{eff}$ is determined by the
in-plane component of the magnetic penetration depth
$\lambda_{ab}$ (see above), $B_{c2}(T)$ has to be measured with
the magnetic field applied parallel to the crystallographic
$c-$axis. The temperature dependence of $B_{c2}^{\parallel c}$
presented in Fig.~\ref{fig:Hc2_corrections}~(a) was obtained from
measurements of the reversible magnetization on the $c-$axis
oriented powder by using the Landau-Ott scaling
approach.\cite{Landau03} The solid line represents fit of
$B_{c2}^{\parallel c}(T)$ using the Werthamer-Helfand-Hohenberg
(WHH) model \cite{Werthamer66} with $B_{c2}^{\parallel
c}(0)=10.7(2)$~T. Note that $B_{c2}^{\parallel c}(0)=10.7(2)$~T
obtained in the present study is close to that reported in the
literature.\cite{Kim02,Zapf05}

The correcting factor $A$  [see
Fig.~\ref{fig:Hc2_corrections}~(b)] was calculated  within the
framework of the numerical Ginzburg-Landau approach developed by
Brandt\cite{Brandt03} in the following way. First, $A$ was
calculated as a function of the reduced field $b=B/B_{c2}$. The
inset in Fig.~\ref{fig:Hc2_corrections}~(b) shows
$A(b)=\sigma\cdot\lambda^{2}$ [see Eq.~(\ref{eq:sigma-lambda})]
for a superconductor with $\kappa=15$, 20, and 25. Vertical lines
correspond to $b=B/B_{c2}(0)$ for the fields used in the present
study (0.05~T, 0.1~T, 0.3~T, and 0.64~T). Second, by using the
calculated WHH form of $B_{c2}(T)$ [red solid line in
Fig.~\ref{fig:Hc2_corrections}~(a)], values of $A(T,B=const)$ for
each particular field $B$ were reconstructed.
Figure~\ref{fig:Hc2_corrections}~(b) shows the calculated $A(T,B)$
at $B=0.05$~T, 0.1~T, and $0.64$~T for $\kappa=15$, 20, and 25.
Note that the influence of $\kappa$ is only important at the
lowest magnetic field ($B=0.05$~T), while at the higher fields the
effect of $\kappa$ becomes almost negligible. This allows us to
estimate the absolute value of $\lambda_{eff}$ at low
temperatures, which was found to be $\lambda_{eff}\simeq120$~nm,
in good agreement with previous results.\cite{Shengelaya05}
Bearing in mind that $\lambda_{eff}\simeq1.31\lambda_{ab}$
\cite{Fesenko91} and $B_{c2}^{\parallel c}=10.7$~T [see
Fig.~\ref{fig:Hc2_corrections}~(a)], the Ginzburg-Landau parameter
was estimated to be $\kappa\simeq17$. This value of $\kappa$ was
used in the following calculations.

\begin{figure}[htb]
\includegraphics[width=1.0\linewidth]{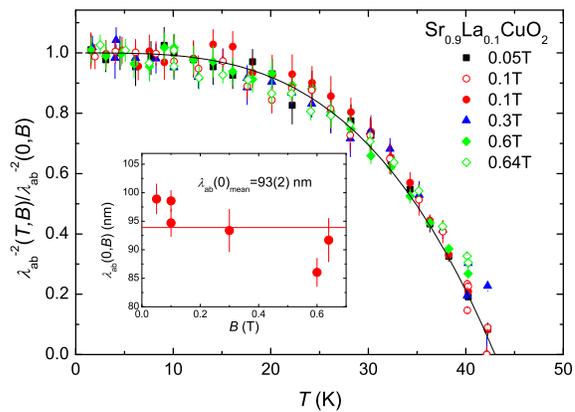}
\caption{(Color online) $\lambda_{ab}^{-2}(T,B)$ of
Sr$_{0.9}$La$_{0.1}$CuO$_{2}$ normalized to their value at $T=0$
reconstructed from $\sigma_{sc}(T)$ measured at $\mu_0H=0.05$~T,
0.1~T, 0.3~T, 0.6~T, and 0.64~T. The solid line is a guide to the
eye. The inset shows $\lambda_{ab}(0)$ as a function of the
applied filed. }
 \label{fig:lambda_normalized}
\end{figure}

The normalized $\lambda_{ab}^{-2}(T,B)/\lambda_{ab}^{-2}(0,B)$
curves reconstructed from the measured $\sigma_{sc}(T,B)$ are
shown in Fig.~\ref{fig:lambda_normalized}. The inset shows
$\lambda_{ab}(0,B)$ obtained from the linear extrapolation of
$\lambda_{ab}^{-2}(T,B)$  at $T<10$~K to zero temperature.
It should be emphasized here that it has no sense to introduce any
unique value of $\lambda$ for each particular magnetic field,
since only the zero-field value of $\lambda$ has a physical
meaning.\cite{Landau07} The same statement holds for the
temperature behavior of $\lambda^{-2}$.  This implies that if the
theory used to reconstruct $\lambda(T)$ from $\sigma_{sc}(T)$
measured in various magnetic fields is correct, than the
corresponding $\lambda(T)$ should {\it coincide}. The data
presented in Fig.~\ref{fig:lambda_normalized} reveal that this is
the case. The normalized
$\lambda_{ab}^{-2}(T,B)/\lambda_{ab}^{-2}(0,B)$ curves merge
together and the value of $\lambda_{ab}$ at $T=0$, presented in
the inset, is almost independent on the magnetic field. The small
increase of $\lambda_{ab}(0)$ at low fields can be explained by
the pinning effects which can lead to a reduction of the second
moment of $P(B)$ in a powder HTS's at fields smaller than $0.1$~T
(see, {\it e.g.}, Refs.~\onlinecite{Zimmermann95} and
\onlinecite{Pumpin90}).

\begin{figure*}[htb]
\includegraphics[width=1.0\linewidth]{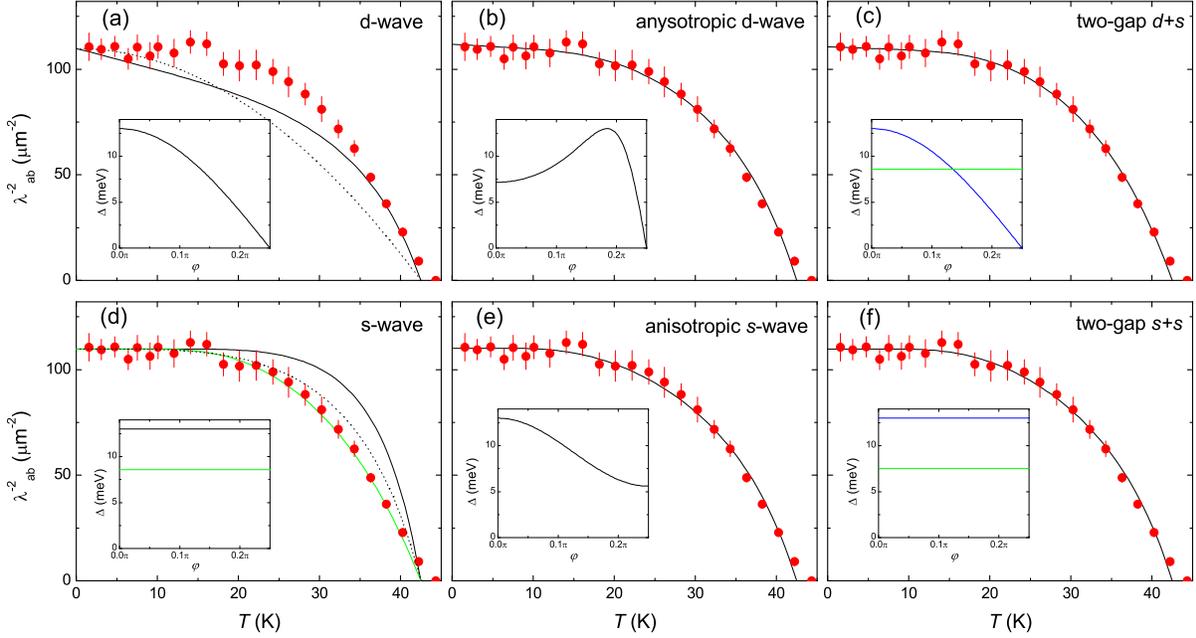}
 \vspace{-1cm}
\caption{(Color online) Temperature dependence of
$\lambda_{ab}^{-2}$ of Sr$_{0.9}$La$_{0.1}$CuO$_{2}$ reconstructed
form $\sigma_{sc} (\mu_0H=0.1$~T). The curves were obtained within
the following models of the gap symmetries: (a) clean $d-$wave
(solid line) and dirty $d-$wave (dotted line), (b) anisotropic
$d-$wave, (c) two-gap $d+s$, (d) clean $s-$wave (solid black --
$\Delta_0=13$~meV and solid green -- $\Delta_0=8.6$~meV) and dirty
$s-$wave (dotted line), (e) anisotropic $s-$wave, (f) two-gap
$s+s$. The corresponding angular dependences of the gaps are shown
in the insets. }
 \label{fig:gap_analysis}
\end{figure*}

Fig.~\ref{fig:gap_analysis} shows $\lambda^{-2}_{ab}(T)$
reconstructed by means of the above described procedure from
$\sigma_{sc}(T)$ for one representative field value $B=0.1$~T. The
data in Fig.~\ref{fig:gap_analysis} were analyzed by using
single-gap and two-gap models, assuming that the superconducting
energy gaps have the following symmetries: $d-$wave (a),
anisotropic $d-$wave (b), $d+s-$wave (c), $s-$wave (d),
anisotropic $s-$wave (e), and $s+s-$wave (f). The analysis for
both the $d-$wave and the isotropic $s-$wave gaps was made in the
clean ($\xi\gg l$, $l$ is the mean-free path) and the dirty
($\xi\ll l$) limit. The cases of anisotropic $s-$  and $d-$wave
gaps, as well as $s+s-$ and $d+s-$wave gap symmetries were
analyzed in the clean limit only.

In the clean limit the temperature dependence of the magnetic
penetration depth $\lambda$ was calculated within the local
(London) approximation ($\lambda\gg\xi$) by using the following
equation:\cite{Tinkham75,Khasanov07_La214}
\begin{equation}
\left.\frac{\lambda^{-2}(T)}{\lambda^{-2}(0)}\right|_{\rm clean}=  1
+\frac{1}{\pi}\int_{0}^{2\pi}\int_{\Delta(T,\varphi)}^{\infty}\left(\frac{\partial
f}{\partial E}\right)\frac{E\
dEd\varphi}{\sqrt{E^2-\Delta(T,\varphi)^2}}.
 \label{eq:lambda-d}
\end{equation}
Here $\lambda^{-2}(0)$ is the zero-temperature value   of the
magnetic penetration depth, $f=[1+\exp(E/k_BT)]^{-1}$ is  the
Fermi function, $\varphi$ is the angle along the Fermi surface
($\varphi=\pi/4$ corresponds to a zone diagonal), and
$\Delta(T,\varphi)=\Delta_0 \tilde{\Delta}(T/T_c)g(\varphi)$
($\Delta_0$ is the maximum gap value at $T=0$).  The temperature
dependence of the gap is approximated by
$\tilde{\Delta}(T/T_c)=\tanh\{1.82[1.018(T_c/T-1)]^{0.51}\}$.\cite{Carrington03}
The function $g(\varphi)$ describes the angular dependence of the
gap and is given by $g^s(\varphi)=1$ for the $s-$wave gap,
$g^d(\varphi)=|\cos(2\varphi)|$ gap for the $d-$wave gap,
$g^{s_{An}}(\varphi)=(1+a\cos4\varphi)/(1+a)$ for the anisotropic
$s-$wave gap,\cite{Shan05} and
$g^{d_{An}}(\varphi)=3\sqrt{3a}\cos2\varphi/2(1+a\cos^2\varphi)^{3/2}$
for the anisotropic $d-$wave gap.\cite{Eremin08}

In the dirty-limit $\lambda^{-2}(T)$ was obtained via:\cite{Tinkham75}
\begin{equation}
\left. \frac{\lambda^{-2}(T)}{\lambda^{-2}(0)}\right|_{{\rm dirty} \ s-{\rm wave}}=
\frac{\Delta(T)}{\Delta(0)}\tanh\left[ \frac{\Delta(T)}{2k_BT}
\right],
 \label{eq:BCS-dirty}
\end{equation}
and assuming the power low dependence
\begin{equation}
\left. \frac{\lambda^{-2}(T)}{\lambda^{-2}(0)}\right|_{{\rm dirty}
\ d-{\rm wave}}=1-\left(\frac{T}{T_c} \right)^n
\end{equation}
with the exponent $n\equiv2$,\cite{Hirschfeld93} for $s-$  and
$d-$wave gaps, respectively.

The two-gap calculations ($d+s-$ and $s+s-$wave) were performed
within the framework of the so called $\alpha-$model assuming that
the total superfluid density is a sum of two
components:\cite{Khasanov07_La214,Carrington03}
\begin{equation}
\frac{\lambda^{-2}(T)}{\lambda^{-2}(0)}=
\omega\cdot\frac{\lambda^{-2}(T,
\Delta_1)}{\lambda^{-2}(0,\Delta_1)}+(1-\omega)\cdot
\frac{\lambda^{-2}(T, \Delta_2)}{\lambda^{-2}(0,\Delta_2)}.
\end{equation}
Here $\Delta_1(0)$ and $\Delta_2(0)$ are the zero-temperature
values of the large and the small gap, respectively, and $\omega$
($0\leq\omega\leq1$) is the weighting factor which represents the
relative contribution of the larger gap to  $\lambda^{-2}$.

The results of the analysis are presented in
Fig.~\ref{fig:gap_analysis}.  The angular dependences of the gaps
[$\Delta_0\cdot g(\varphi)$] are shown in the corresponding
insets. The maximum value of the gap $\Delta_0=13$~meV was kept
fixed in accordance with the results of tunneling
experiments.\cite{Chen02} It is obvious that simple $d-$ and
$s-$wave approaches with $\Delta_0=13$~meV cannot describe the
observed $\lambda^{-2}_{ab}(T)$ [see
Figs.~\ref{fig:gap_analysis}~(a) and (d)]. All other models as,
{\it e.g.}, both anisotropic $d-$ and $s-$wave, and both two-gap
$d+s$ and $s+s$ [Figs.~\ref{fig:gap_analysis}~(b), (c), (e), (f)],
as well as the single $s-$wave model with $\Delta_0=8.6$~meV
[solid green line in Fig.~\ref{fig:gap_analysis}~(a)] describe the
experimental data reasonably well. The results of the analysis are
summarized in Table~\ref{Table:results}.

\begin{table}[h]
\caption[~]{\label{Table:results} Summary of the gap analysis of
the temperature dependences of $\lambda^{-2}_{ab}$ for
Sr$_{0.9}$La$_{0.1}$CuO$_{2}$.  The meaning of the parameters is
explained in the text.}
\begin{center}
 \vspace{-0.7cm}
\begin{tabular}{lcccccccc}\\ \hline
\hline
 Model&$\Delta_{0,1}$&$a$&$\Delta_{0,2}$&$\omega$\\
&(meV)&&(meV)&\\
\hline
Clean limit $s$&8.6&--&--&--\\
Anisotropic $d$&13\footnotemark[1]&3.1&--&--\\
Anisotropic $s$&13\footnotemark[1]&0.4&--&--\\
Two-gap $d+s$&13\footnotemark[1]&--&9.0&0.15\\
Two-gap $s+s$&13\footnotemark[1]&--&7.5&0.35\\
 \hline
 \hline
\end{tabular}
 \footnotetext[1]{\ From tunnelling experiments Ref.~\onlinecite{Chen02}}
   \end{center}
\end{table}

\begin{figure}[htb]
\includegraphics[width=0.9\linewidth]{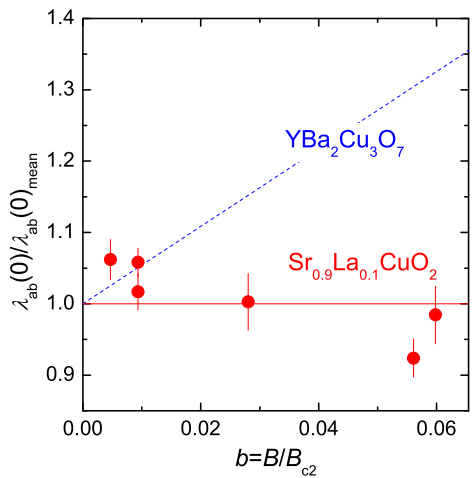}
 \vspace{-0.7cm}
\caption{(Color online) $\lambda_{ab}(0)$ of
Sr$_{0.9}$La$_{0.1}$CuO$_{2}$ normalized to its mean value as a
function of reduced magnetic filed $b=B/B_{c2}$. The dashed blue
line line corresponds to $\lambda_{ab}(0,b)$ for
YBa$_2$Cu$_3$O$_{7-\delta}$ from Ref.~\onlinecite{Sonier00}.}
 \label{fig:lambda_field-dependence}
\end{figure}

Based on our analysis we cannot differentiate between the models
presented in Figs.~\ref{fig:gap_analysis}~(b), (c), (e), (f), and
the isotropic $s-$wave model with $\Delta_0=8.6$~meV [solid green
line in Fig.~\ref{fig:gap_analysis}~(d)]. We believe, however,
that from our consideration we can exclude the case of an
anisotropic $d-$wave gap [see Fig.~\ref{fig:gap_analysis}~(b)].
The reasons are the following: (i) The model of Brandt,
\cite{Brandt03} used to reconstruct $\lambda^{-2}_{ab}(T)$ from
measured $\sigma_{sc}(T)$, is strictly valid for conventional
superconductors only. The presence of nodes in the gap makes the
electrodynamics of the mixed state highly nonlocal, leading to an
additional source for decreasing the superfluid density due to
excitation of the quasiparticles along the nodal directions (see,
{\it e.g.}, Ref.~\onlinecite{Kadono04} and references therein). In
this case the proportionality coefficient $A$ [see
Eq.~(\ref{eq:sigma-lambda}) and
Fig.~\ref{fig:Hc2_corrections}~(b)] depends much stronger on the
reduced magnetic field and, as a consequence, on temperature than
it is expected in a case of conventional superconductors. However,
the normalized $\lambda^{-2}_{ab}(T,B)/\lambda^{-2}_{ab}(0,B)$
evaluated at various fields merge together (see
Fig.~\ref{fig:lambda_normalized}), implying that
Sr$_{0.9}$La$_{0.1}$CuO$_{2}$ can be, indeed, described within the
model developed for conventional isotropic
superconductors.\cite{Brandt03} (ii) The nonlocal and the
nonlinear response of the superconductor with nodes in the gap to
the magnetic field leads to the fact that $\lambda_{ab}(0)$,
evaluated from $\mu$SR experiments, becomes magnetic field
dependent and increases with increasing field.\cite{Amin00} Such a
behavior was observed in various hole-doped
HTS's.\cite{Kadono04,Sonier00,Khasanov07_field-effect} For
comparison, in Fig.~\ref{fig:lambda_field-dependence} we show
$\lambda(0,b)$ for Sr$_{0.9}$La$_{0.1}$CuO$_{2}$ measured here and
the one obtained by Sonier {\it et al.}\cite{Sonier00} for
hole-doped YBa$_2$Cu$_3$O$_{7-\delta}$. Whereas $\lambda_{ab}(0)$
for YBa$_2$Cu$_3$O$_{7-\delta}$ strongly increases with magnetic
field, for Sr$_{0.9}$La$_{0.1}$CuO$_{2}$ it is almost field
independent.

\section{Summary and Conclusions}

Muon-spin rotation measurements  were performed on the
electron-doped high-temperature cuprate superconductor
Sr$_{0.9}$La$_{0.1}$CuO$_{2}$ ($T_c\simeq43$~K).
$\lambda_{ab}^{-2}(T)$ was reconstructed from the measured
temperature dependence of the $\mu$SR linewidth by using
numerical calculations of Brandt.\cite{Brandt03} The main results
are:
(i) The absolute value of the in-plane magnetic penetration depth
$\lambda_{ab}$ at $T=0$ was found to be
$\lambda_{ab}(0)=93(2)$~nm.
(ii) $\lambda_{ab}$ is independent of the magnetic field in
contrast to the strong magnetic field dependence observed in
hole-doped HTS's. This suggests that there are no nodes in the
superconducting energy gap of Sr$_{0.9}$La$_{0.1}$CuO$_{2}$.
(iii) The temperature dependence of $\lambda_{ab}^{-2}$  was found
to be inconsistent with an isotropic $s-$wave as well as with a
$d-$wave symmetry of the superconducting energy gap, in both the
clean and the dirty limit.
(iv) $\lambda_{ab}^{-2}(T)$ is well described by anisotropic
$s-$wave and two-gap models ($d+s$ and $s+s$). In the case of the
two-gap $d+s$ model the contribution of the $d-$wave gap to the
total superfluid density was estimated to be $\simeq$15\%.
To conclude, a substantial $s-$wave component in the
superconducting order parameter is present in
Sr$_{0.9}$La$_{0.1}$CuO$_{2}$, in agreement with previously
reported results of tunneling,\cite{Chen02} specific
heat,\cite{Liu05} and small-angle neutron scattering
experiments.\cite{White08}
%

\section{Acknowledgments}

This work was partly performed at the Swiss Muon Source (S$\mu$S),
Paul Scherrer Institute (PSI, Switzerland). The authors are
grateful to A.~Amato and D.~Herlach for providing the instrumental
support,  and  D.G.~Eshchenko for help during the $\mu$SR
experiments. This work was supported by the Swiss National Science
Foundation, by the K.~Alex M\"uller Foundation and in part by the
SCOPES grant No. IB7420-110784 and the EU Project CoMePhS.


\begin{thebibliography}{99}
%
\bibitem{Tsuei94} C.C.~Tsuei,  J.R.~Kirtley, C.C.~Chi, L.S.~Yu-Jahnes,
A.~Gupta, T.~Shaw, J.Z.~Sun, and M.B.~Ketchen, Phys.~Rev.~Lett. {\bf 73}, 593 (1994).
%
\bibitem{Khasanov07_La214} R.~Khasanov, A.~Shengelaya, A.~Maisuradze, F.~La~Mattina,
A.~Bussmann-Holder, H.~Keller, and K.A.~M\"uller, Phys.~Rev.~Lett.
{\bf 98}, 057007 (2007).
%
\bibitem{Khasanov07_Y124} R.~Khasanov, A.~Shengelaya,
A.~Bussmann-Holder, J.~Karpinski, H.~Keller, and  K.A.~M\"uller,
J.~Supercond.~Nov.~Magn. {\bf 21}, 81 (2008).
%
\bibitem{Khasanov07_Y123} R.~Khasanov, S.~Str\"assle, D.~Di~Castro,
T.~Masui, S.~Miyasaka, S.~Tajima, A.~Bussmann-Holder, and
H.~Keller, Phys.~Rev.~Lett. {\bf 99}, 237601 (2007).
%
\bibitem{Sato01} T.~Sato,  T.~Kamiyama,  T.~Takahashi,
K.~Kurahashi, and K.~Yamada, Science {\bf 291}, 1517 (2001).
%
\bibitem{Matsui05} H.~Matsui, K.~Terashima, T.~Sato,
T.~Takahashi, M.~Fujita, and K.~Yamada,  Phys.~Rev.~Lett. {\bf
95}, 017003 (2005).
%
\bibitem{Tsuei00} C.C.~Tsuei and J.R.~Kirtley,
Phys.~Rev.~Lett. {\bf 85}, 182
(2000).
%
\bibitem{Blumberg02} G.~Blumberg, A.~Koitzsch,
A.~Gozar, B.S.~Dennis, C.A.~Kendziora, P.~Fournier,  and
R.L.~Greene, Phys.~Rev.~Lett. {\bf 88}, 107002 (2002).
%
\bibitem{Kokales00}  J.D.~Kokales, P.~Fournier, L.V.~Mercaldo,
V.V.~Talanov, R.L.~Greene, and S.M.~Anlage,  Phys.~Rev.~Lett. {\bf
85}, 3696 (2000).
%
\bibitem{Snezhko04} A.~Snezhko, R.~Prozorov, D.D.~Lawrie, R.W.~Giannetta, J.~Gauthier, J.~Renaud, and P.~Fournier, Phys.~Rev.~Lett. {\bf 92}, 157005 (2004).
%
\bibitem{Shan05} L.~Shan, Y.~Huang, H.~Gao, Y.~Wang, S.L.~Li, P.C.~Dai, F.~Zhou, J.W.~Xiong, W.X.~Ti, and H.H.~Wen, Phys.~Rev.~B {\bf 72} 144506 (2005).
%
\bibitem{Chen02} C.-T.~Chen, P.~Seneor, N.-C.~Yeh, R.P.~Vasquez, L.D.~Bell, C.U.~Jung,  J.Y.~Kim, M.-S.~Park, H.-J.~Kim, and S.-Ik.~Lee, Phys.~Rev.~Lett. {\bf 88}, 227002 (2002).
%
\bibitem{Hackl95-03} B.~Stadlober, G.~Krug, R.~Nemetschek, R.~Hackl, J.L.~Cobb, and J.T.~Markert, Phys.~Rev.~Lett. {\bf 74}, 4911 (1995); F.~Venturini, R. Hackl, and U.~Michelucci, Phys.~Rev.~Lett. {\bf 90}, 149701 (2003).
%
\bibitem{Alff99} L.~Alff, S.~Meyer, S.~Kleefisch, U.~Schoop, A.~Marx, H.~Sato, M.~Naito, and R.~Gross, Phys.~Rev.~Lett. {\bf 83}, 2644 (1999).
%
\bibitem{White08} J.S.~White, E.M.~Forgan, M.~Laver,
P.S.~H\"afliger,  R.~Khasanov, R.~Cubitt, C.D.~Dewhurst,
M.S.~Park, D.-J.~Jang, and S.-I.~Lee, J.~Phys.:~Condens.~Matter.
{\bf 20}, 104237 (2008).
%
\bibitem{Liu05} Z.Y.~Liu, H.H.~Wen, L.~Shan, H.P~Yang, X.F.~Lu, H.~Gao,
M.-S.~Park, C.U.~Jung, and S.-I.~Lee,  Europhys.~Lett. {\bf 69}, 263 (2005).
%
\bibitem{Biswas02} A.~Biswas, P.~Fournier, M.M.~Qazilbash, V.N.~Smolyaninova, H.~Balci, and R.L.~Greene, Phys.~Rev.~Lett. {\bf 88}, 207004 (2002).
%
\bibitem{Skinta02} J.A.~Skinta, M.-S.~Kim, T.R.~Lemberger, T.~Greibe, and M.~Naito, Phys.~Rev.~Lett. {\bf 88}, 207005 (2002).
%
\bibitem{Luo05} H.G.~Luo and T.~Xiang, Phys.~Rev.~Lett. {\bf 94}, 027001 (2005).
%
\bibitem{Liu06} C.S.~Liu, H.G.~Luo, W.C.~Wu, and T.~Xiang, Phys.~Rev.~B {\bf 73},
174517 (2006).
%
\bibitem{Siegrist88} T.~Siegrist, S.M.~Zahurak, D.W.~Murphy, and R.S.~Roth, Nature(London) {\bf 334}, 231 (1988).
%
\bibitem{Smith91} M.G.~Smith, A.~Manthiram, J.~Zhou, J.B.~Goodenough, and  J.T.~Markert, Nature(London) {\bf 351} 549 (1991).
%
\bibitem{Jorgensen93} J.D.~Jorgensen, P.G.~Radaelli, D.G.~Hinks, J.L.~Wagner, S.~Kikkawa, G.~Er, and F.~Kanamaru, Phys.~Rev.~B {\bf 47}, 14654 (1993).
%
\bibitem{Brandt03} E.H.~Brandt, Phys.~Rev.~B {\bf 68}, 054506
(2003).
%
\bibitem{Jung02} C.U.~Jung, J.Y.~Kim, M.-S.~Kim, M.-S.~Park, H.-J.~Kim, Y.~Yao, S.Y.~Lee, and S.-I.~Lee, Physica~C {\bf 366}, 299 (2002).
%
\bibitem{Brandt88} E.H.~Brandt, Phys.~Rev.~B {\bf 37}, 2349
(1988).
%
\bibitem{Khasanov06a} R.~Khasanov, I.L.~Landau, C.~Baines, F.~La~Mattina, A.~Maisuradze, K.~Togano, and H.~Keller, Phys.~Rev.~B {\bf 73}, 214528 (2006).
%
\bibitem{Zimmermann95} P.~Zimmermann, H.~Keller,
S.L.~Lee, I.M.~Savi\'c, M.~Warden, D.~Zech, R.~Cubitt,
E.M.~Forgan, E.~Kaldis, J.~Karpinski, and C.~Kr\"uger,
Phys.~Rev.~B {\bf 52},  541  (1995).
%
\bibitem{Fesenko91}  V.I.~Fesenko, V.N.~Gorbunov, and V.P.~Smilga, Physica~C {\bf 176}, 551 (1991).
%
\bibitem{Landau03} I.L.~Landau and H.R.~Ott, Physica~C {\bf 385}, 544 (2003).
%
\bibitem{Werthamer66}  N.R.~Werthamer, E.~Helfand, and P.C.~Hohenberg, Phys.~Rev. {\bf 147}, 295 (1966).
%
\bibitem{Kim02} M.-S.~Kim, T.R.~Lemberger, C.U.~Jung, J.-H.~Choi, J.Y.~Kim, H.-J.~Kim, and S.-Ik.~Lee, Phys.~Rev.~B {\bf 66}, 214509 (2002).
%
\bibitem{Zapf05} V.S.~Zapf, N.-C.~Yeh, A.D.~Beyer, C.R.~Hughes, C.H.~Mielke, N.~Harrison, M.S.~Park, K.H.~Kim, and S.-I.~Lee, Phys.~Rev.~B {\bf 71}, 134526 (2005).
%
\bibitem{Shengelaya05} A.~Shengelaya, R.~Khasanov, D.G.~Eshchenko, D.~Di~Castro, I.M.~Savi\'c, M.S.~Park, K.H.~Kim, S.-I.~Lee, K.A.~M\"uller, and H.~Keller, Phys.~Rev.~Lett. {\bf 94}, 127001 (2005).
%
\bibitem{Landau07} I.L.~Landau and  H.~Keller, Physica~C {\bf 466}, 131 (2007).
%
\bibitem{Pumpin90} B.~P\"umpin, H.~Keller, W.~K\"undig, W.~Odermatt, I.M.~Savi\'c, J.W.~Schneider, H.~Simmler, P.~Zimmermann, E.~Kaldis, S.~Rusiecki, Y.~Maeno, and C.~Rossel, Phys.~Rev.~B {\bf 42}, 8019 (1990).
%
\bibitem{Tinkham75} M.~Tinkham, ''Introduction to Superconductivity``, {\it Krieger Publishing company, Malabar, Florida, 1975}.
%
\bibitem{Carrington03} A.~Carrington and F.~Manzano, Physica~C {\bf 385}, 205 (2003).
%
\bibitem{Eremin08} I.~Eremin, E.~Tsoncheva, and A.V.~Chubukov, Phys.~Rev.~B {\bf 77}, 024508 (2008).
%
\bibitem{Hirschfeld93} P.J.~Hirschfeld and N.~Goldenfeld,
Phys.~Rev.~B {\bf 48}, 4219 (1993).
%
\bibitem{Kadono04} R.~Kadono, J.~Phys.:~Condens.~Matter {\bf 16}, S4421 (2004).
%
\bibitem{Amin00} M.H.S.~Amin, M.~Franz, and I.~Affleck, Phys.~Rev.~Lett. {\bf 84},
5864 (2000).
%
\bibitem{Sonier00} J.E.~Sonier, J.H.~Brewer, and R.F.~Kiefl, Rev.~Mod.~Phys. {\bf 72}, 769 (2000).
%
\bibitem{Khasanov07_field-effect}  R.~Khasanov, A.~Shengelaya, D.~Di~Castro, D.G.~Eshchenko,
I.M.~Savi\'c, K.~Conder, E.~Pomjakushina, J.~Karpinski,
S.~Kazakov, and H.~Keller, Phys.~Rev.~B {\bf 75}, 060505 (2007).
%
\end{thebibliography}
\end{document}